\documentclass[aip,jcp,reprint,noshowkeys,superscriptaddress]{revtex4-1}
\usepackage{subcaption}
\usepackage{bm,graphicx,tabularx,array,booktabs,dcolumn,xcolor,microtype,multirow,amscd,amsmath,amssymb,amsfonts,physics,siunitx,upgreek}
\usepackage[version=4]{mhchem}
\usepackage[utf8]{inputenc}
\usepackage[T1]{fontenc}
\usepackage{txfonts}
\usepackage[normalem]{ulem}
\usepackage{mleftright}
\usepackage{gensymb}

\usepackage[margin=1.8cm]{geometry}
\usepackage{lipsum}

\usepackage{tikz}
\usetikzlibrary{arrows.meta,positioning,shapes.misc,arrows}

\usepackage[
	colorlinks=true,
    citecolor=blue,
    linkcolor=blue,
    filecolor=blue,      
    urlcolor=blue,	
    breaklinks=true
	]{hyperref}
\usepackage{titlesec}
\titlespacing\section{10pt plus 2pt minus 2pt}{10pt plus 2pt minus 2pt}{10pt plus 2pt minus 2pt}
\titlespacing\subsection{2pt plus 2pt minus 2pt}{10pt plus 2pt minus 2pt}{5pt plus 2pt minus 2pt}
\titlespacing\subsubsection{2pt plus 2pt minus 2pt}{10pt plus 2pt minus 2pt}{5pt plus 2pt minus 2pt}

\usepackage{caption} 
\usepackage{subcaption}

\usepackage{natbib}
\AtBeginDocument{\nocite{achemso-control}}

\newcommand{\sigg}{\upsigma_\text{g}}
\newcommand{\sigu}{\upsigma_\text{u}}

\newcommand{\Eh}{\mathrm{E_h}}

\newcommand{\cQ}{\hat{\mathcal{Q}}}

\newcommand{\hI}{\hat{I}}
\newcommand{\hH}{\hat{H}}
\newcommand{\hV}{\hat{V}}

\newcommand{\Ne}{N_\text{e}}

\newcommand{\br}{\bm{r}}

\newcommand{\lep}{\lambda_\text{c}}
\newcommand{\rc}{r_\text{c}}

\newcommand{\etal}{\textit{et al.}}

\definecolor{hughgreen}{RGB}{102, 0, 204}

\newcommand{\UOX}{Physical and Theoretical Chemistry Laboratory, Department of Chemistry, University of Oxford, South Parks Road, Oxford, OX1 3QZ, U.K.}
\newcommand{\MPICPS}{Max Planck Institute for Chemical Physics of Solids, 01187 Dresden, Germany}

\begin{document}

\title{Convergence of M\o{}ller--Plesset perturbation theory for excited reference states}
\author{Diana-Gabriela Oprea}
\affiliation{\UOX}
\affiliation{Current address: \MPICPS}

\author{Hugh~G.~A.~Burton}
\email{hgaburton@gmail.com}
\affiliation{\UOX}

\date{\today}

\begin{abstract}
Excited states in molecules can be difficult to investigate and 
generally require methods that are either computationally expensive or are not universally accurate.
Recent research has focused on using higher-energy Slater determinants as mean-field representations
of excited states, which can then be used to define reference states for electron correlation techniques such as
M\o{}ller--Plesset theory.
However, the convergence behaviour of these excited-state perturbation series has not yet been 
explored, limiting our understanding into the systematic improvability and reliability of such methods.
Here, we present a systematic analysis of the M\o{}ller--Plesset perturbation series 
for closed-shell excited states using higher-energy reference determinants
identified with, and without, state-specific orbital optimisation.
We combine numerical calculations with complex analysis to show that the excited-state M\o{}ller--Plesset
theory diverges almost universally for the stereotypical \ce{H2}, \ce{H2O}, and \ce{CH2} molecules.
Our results cast doubt on the suitability of higher-energy Slater determinants as reference states
for perturbative treatments of electron correlation.
\end{abstract}

\maketitle
\raggedbottom
\linepenalty1000

\section{Introduction}
Excited states are essential for understanding photochemical processes, ranging from photosynthesis and 
biological mechanisms of vision, to photocatalytic reactions and optoelectronics.\cite{GonzalezBook}
Single reference \textit{ab initio} methods have been extensively developed to predict these excited states, including 
time-dependent density functional theory\cite{Runge1984,Dreuw2005,Burke2005} (TD-DFT), 
equation-of-motion coupled-cluster\cite{Stanton1993,Krylov2008} (EOM-CC), or the 
algebraic diagrammatic construction\cite{Schirmer1982,Dreuw2015} (ADC) approach.
While TD-DFT works well for single excitations, it cannot necessarily describe double 
excitations or the orbital relaxation that occurs in charge transfer processes.\cite{Tozer2003,Dreuw2004}
Furthermore, describing double excitations with the EOM-CC and ADC approach requires high
excitation orders that increase the computational scaling.
The multiconfigurational complete active space self-consistent field\cite{Roos1980a,Roos1980b} approach
with second-order perturbation theory\cite{Andersson1990,Andersson1992} (CASPT2) can provide more accurate results, but 
has a computational cost that grows exponentially with the active space.

To overcome these challenges, the idea of describing excitations using higher-energy stationary 
points that are present in wave function methods has seen renewed interest.
These excited state solutions correspond to local minima,  saddle points, or local maxima of 
the parameterised electronic energy landscape.\cite{Burton2021a,Burton2022a}
In particular, it has been recognised that higher-energy self-consistent field (SCF) solutions from Hartree--Fock (HF)
theory or Kohn--Sham DFT can provide qualitative representations of double excitations, and can incorporate  
essential orbital relaxation effects.\cite{Gilbert2008,Barca2018}
The development of optimisation algorithms to prevent variational collapse to the ground state 
now enables routine calculations to be performed using this state-specific 
philosophy.\cite{Gilbert2008,Ye2017,Ye2019,Hait2021,CarterFenk2020,Levi2020,Levi2020a,Ivanov2021,Schmerwitz2022}

While excited-state SCF stationary points can qualitatively describe double excitations, they do 
not capture any of the dynamic correlation that is generally required for quantitative accuracy.
Therefore, it has been suggested that these higher-energy Slater determinants can be used as a reference 
state for standard post-HF correlation techniques, including M\o{}ller--Plesset (MP)
perturbation theory\cite{Warken1995,Lee2019b} and 
coupled-cluster (CC) theory.\cite{Meissner1993,Jankowski1994,Jankowski1994a,Kowalski1998,Marie2021a,Kossoski2021}
Recent studies have demonstrated that excited state-specific CC ($\Delta$CC) calculations can give very accurate 
energies, but can also exhibit multiple solutions representing each excitation and 
amplitude equations that are difficult to converge.\cite{Mayhall2010} 
On the other hand, excited-state second-order MP perturbation theory ($\Delta$MP2) appears to provide relatively
accurate results for double excitations, particularly when an orbitally-optimised reference determinant 
is employed.\cite{Lee2019b}

In the absence of exact results, electronic structure theory benefits from a systematically
improvable hierarchy of methods that allows the error of low-order approximations to be reliably assessed.
Therefore, the MP$n$ perturbation series should ideally be convergent.
Satisfying this condition justifies the use of low-order approximations, such as second-order MP2, and provides confidence
that the corresponding energies are reliable.
Since the series convergence clearly cannot be tested on each occasion, we must rely 
on benchmark studies to understand which chemical scenarios correspond to convergent 
perturbation expansions.
Extensive research, reviewed in Ref.~\onlinecite{Marie2021}, has shown that ground-state perturbation theory
converges when only one configuration is dominant and there are no low-lying 
excited states.\cite{Gill1988}
On the other hand, divergent series are created by interactions with nearly degenerate 
states (known as intruder states),\cite{Christiansen1996,Olsen2000,Olsen2019}
or critical points associated with dense 
electron clusters in systems like \ce{Ne}, \ce{HF}, or \ce{F-}.\cite{Olsen1996,Cremer1996,Stillinger2000,Goodson2004,Sergeev2005}
Furthermore, spin-contamination in unrestricted reference states\cite{SzaboBook}
 is known to give particularly 
slow convergence for the corresponding unrestricted MP series.\cite{Laidig1985,Knowles1985,Handy1985,Gill1988}


Excited states do not appear to satisfy these general conditions for a convergent MP series.
They often have open-shell multi-configurational character 
and occupy dense regions of the energy spectrum with many near degeneracies.
Therefore, we expect the issues associated with divergent ground-state perturbation theory to be 
even more prevalent, casting doubt on the general suitability of excited-state 
perturbation methods such as $\Delta$MP2. 
In this work, we test this hypothesis by performing a systematic analysis 
of the MP series convergence for excited states in stereotypical small molecules
using a combination of complex analysis and numerical calculations.
We investigate how the convergence depends on the molecular geometry and the 
inclusion of orbital relaxation in the reference determinant.
To avoid spin contamination and degenerate open-shell reference states, we limit ourselves to
the double excitations in \ce{H2}, \ce{H2O}, and \ce{CH2} using a minimal basis set.

\section{Theory}
\label{sec:pt_theory}
\subsection{M\o{}ller--Plesset perturbation theory}
\label{subsec:MPtheory}
Rayleigh--Schr\"{o}dinger perturbation theory proceeds by recasting the time-independent
Schr\"{o}dinger equation as
\begin{equation}
\hH(\lambda) \Psi_k(\lambda) = (\hH^{(0)} + \lambda \hV) \Psi_k(\lambda) = E_k(\lambda) \Psi(\lambda),
\label{eq:tise}
\end{equation}
where $\hH^{(0)}$ is a zeroth-order Hamiltonian with known eigenfunctions $\Psi_k^{(0)}$ satisfying
\begin{equation}
\hH^{(0)}\, \Psi_k^{(0)} = E_k^{(0)}\, \Psi_k^{(0)}, 
\end{equation}
and $\hV = \hH - \hH^{(0)}$ represents the perturbation.
These reference solutions correspond to the eigenstates of $\hH(\lambda)$ with $\lambda = 0$ while 
the physically exact states arise for $\lambda = 1$.
Assuming a small perturbation, the wave function and energy can be expanded as an 
infinite power series around $\lambda = 0$ to give
\begin{subequations}
\begin{align}
E_k    &= \sum_{j=0}^{\infty} E_k^{(j)} \lambda^j,
\label{eq:Eseries}
\\
\Psi_k &= \sum_{j=0}^{\infty} \Psi_k^{(j)} \lambda^j,
\end{align}
\end{subequations}
where $E_k^{(j)}$ and $\Phi_k^{(j)}$ are the $j^\text{th}$ order corrections to the wave function and energy, 
respectively, and are formally given by the derivatives
\begin{equation}
E_k^{(j)} = \frac{1}{j!} \left. \frac{\partial^{j}\, E_k(\lambda)}{\partial \lambda^j} \right\vert_{\lambda = 0} 
\quad\text{and}\quad
\Psi_k^{(j)} = \frac{1}{j!} \left. \frac{\partial^{j}\, \Psi_k(\lambda)}{\partial \lambda^j} \right\vert_{\lambda = 0}. 
\end{equation}
In practice, the wave function and energy corrections can be obtained using the iterative expressions\cite{HelgakerBook}
\begin{subequations}
\begin{align}
E^{(j)}_k    &= \mel*{\Phi^{(0)}_k}{\hV}{\Phi^{(j-1)}_k}
\\
\Psi^{(j)}_k &= \cQ\,\qty(\hH^{(0)} - E^{(0)}_k)^{-1} \cQ\, \qty( - \hV\, \Psi^{(j-1)}_k + \sum_{i=1}^j E^{(i)}_k \Psi^{(j-i)}_k),
\label{eqn:gen_energy}
\end{align}
\end{subequations}
where $\cQ = \hI - \ket*{\Psi_k^{(0)}}\bra*{\Psi_k^{(0)}}$ projects out the reference state 
to ensure that the operator $(\hH^{(0)} - E^{(0)}_k)$ is invertible (for a non-degenerate $\Psi_k^{(0)}$).
 

The most common electronic perturbation expansion is M\o{}ller--Plesset (MP) theory.\cite{Moller1934}
Within MP theory, the unperturbed Hamiltonian for an $\Ne$ electron system 
is defined as the sum of one-particle Fock operators 
\begin{equation}
\label{eqn:h_fock}
\hH^{(0)} = \sum_{i=1}^{\Ne} \hat{F}(\br_i)
\end{equation}
and the reference state corresponds to a single Slater determinant.\cite{SzaboBook}
The Fock operator and reference determinant are usually identified as a mean-field solution 
to the HF equations, although the use of alternative orbitals 
can improve the accuracy in certain cases.\cite{Bertels2019,Rettig2020}
If the reference state is a SCF solution of the Fock operator, then 
every possible Slater determinant is an eigenstate of $\hH^{(0)}$ with the eigenvalue 
corresponding to the sum of the occupied orbital energies.

The MP$n$ approximant truncates Eq.~\eqref{eq:Eseries} at $n^\text{th}$ order, giving
\begin{equation}
E_k^{n} = \sum_{j=0}^{n} E_k^{(j)} 
\end{equation}
where $\lambda = 1$ has been applied to obtain an estimate of the physical energy.
The MP1 approximant corresponds to the total energy of the reference state.
In practice, most calculations stop at the MP2 level, which gives the lowest-order correction to the 
energy and is generally more accurate than the MP3 approximation.
Furthermore, standard MP2 implementations retain a relatively low $\mathcal{O}(N^5)$ computational scaling, 
arising from the two-electron integral transformation, while lower $\mathcal{O}(N^4)$ scaling can be achieved 
by working directly in the atomic orbital basis.\cite{Haser1992,Ayala1999,Surjan2005}

\subsection{Choice of reference state and Hamiltonian partitioning}
\label{subsec:ref_part}

In principle, the MP perturbation expansion can be applied to 
any ground or excited state $\Psi_k$ using a suitable reference wave function and Hamiltonian.
There are two options for targeting excited states with single-determinant approximations.
The first is $\Delta$SCF, which constructs an excited-state Slater determinant by occupying the relevant 
higher-energy orbitals obtained from a ground-state calculation. 
Alternatively, the $\Delta$SCF orbitals can be used as a starting point for a 
higher-energy state-specific HF calculation.
Algorithms such as the Maximum Overlap Method\cite{Gilbert2008} (MOM) 
prevent variational collapse to the ground state by altering the orbital selection step in the
self-consistent field approach. 
Instead of selecting the orbitals with the lowest energy, the orbitals are chosen as those that 
have the largest overlap with the occupied orbitals on the previous iteration.
As long as a suitable initial guess can be found, this strategy self-consistently optimises 
the orbitals such that the final determinant corresponds to 
a higher-energy stationary point on the HF energy landscape.\cite{Burton2021a} 
Therefore, we might expect that MOM-SCF will give a more accurate reference for 
excited-state perturbation theory.

Choosing $\Delta$SCF or MOM-SCF affects the reference Fock operator in addition to 
the reference wave function and orbitals
For $\Delta$SCF, the zeroth-order Hamiltonian corresponds to the ground-state Fock operator such 
that the excited Slater determinant is an eigenfunction of the reference problem.
In contrast, the reference Hamiltonian for MOM-SCF is the optimal Fock matrix 
identified in the higher-energy HF calculation.
This redefinition of the Fock matrix, and thus the orbital energies, has the potential to change
the convergence behaviour of the perturbation series in an unpredictable manner.


\subsection{Assessing the convergence using complex analysis}
\label{subsec:complex}
While the eigenvalues of $\hH(\lambda)$ at $\lambda = 1$ must correspond to the exact ground and 
excited state energies, there is no guarantee that the perturbation series in 
Eq.~\eqref{eq:Eseries} is convergent.
Divergent expansions mean that the energy approximants $E_k^n$ become increasingly inaccurate
for large $n$. 
Therefore, the corresponding perturbation series is not systematically improvable and it is difficult to 
assess the reliability of the low-order approximations that are generally applied in practice (e.g.\ MP2).


\begin{figure*}[t!]
\includegraphics[width=\linewidth]{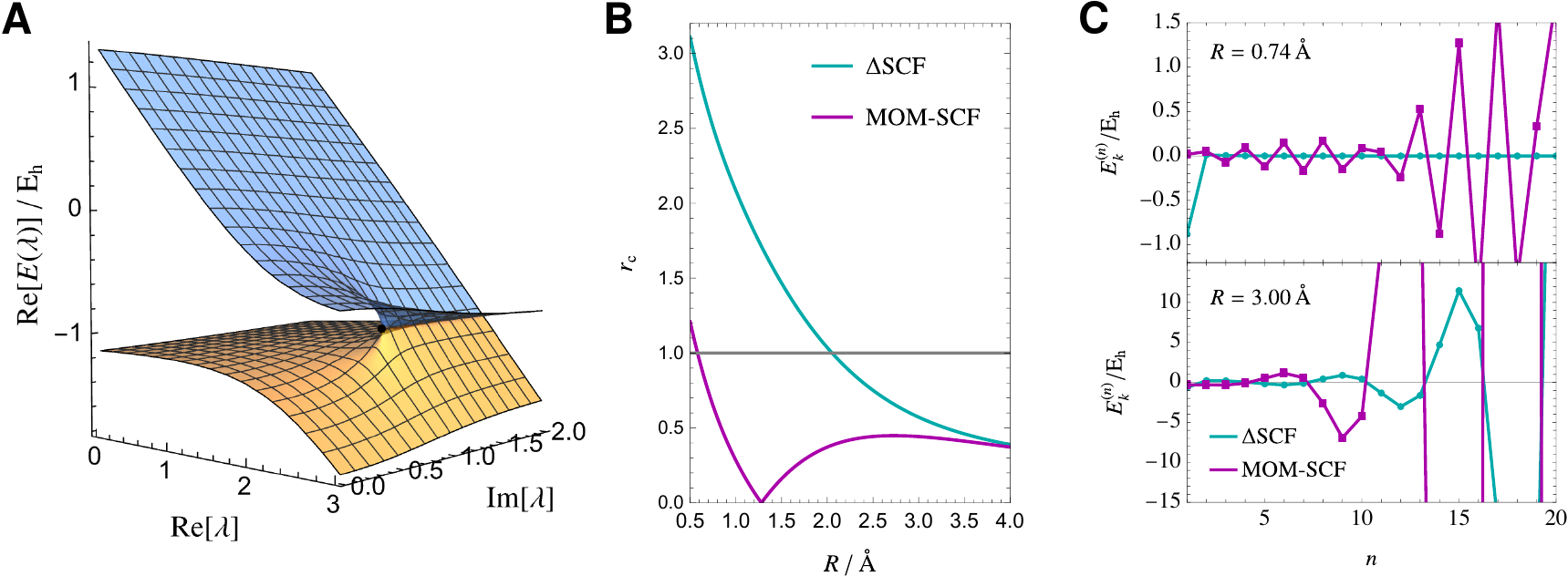}
\caption{ (\textbf{\textsf{A}}) The ground and excited eigenstates of $\hH(\lambda)$ in the complex-$\lambda$ plane are connected at a square-root exceptional point, shown here for the 
$\Delta$SCF reference Hamiltonian at $R(\ce{H-H}) = \SI{0.76}{\angstrom}$.
(\textbf{\textsf{B}}) Comparison of the radius of convergence for the $\Delta$SCF and MOM-SCF strategies.
(\textbf{\textsf{C}}) The MP$n$ corrections for the $\sigu^2$ excited state of \ce{H2} at the 
equilibrium and stretched bond lengths.
}
\label{fig:h2convergence}
\end{figure*}

The radius of convergence $\rc$ defines the values of $\lambda$ for which the expansion~\eqref{eq:Eseries}
gives a convergent series. 
In particular, a convergent series occurs when $\abs{\lambda} < \rc$.
Mathematically, the radius of convergence is determined by the distance of the closest singularity of the 
energy $E_k(\lambda)$ from the origin in the complex-$\lambda$ plane, known as the `dominant' singularity $\lep$.\cite{Marie2021}
This singularity corresponds to a point where the energy function $E_k(\lambda)$ is non-analytic such that 
the function is not infinitely differentiable with respect to $\lambda$. 
These points may correspond to a pole or a branch point (exceptional point) of a complex-valued function.
Therefore, we require $\rc > 1$ for the perturbation expansion to be convergent at the physically relevant
value $\lambda = 1$.

In the context of perturbation theory, the energy surfaces $E_k(\lambda)$ correspond to the eigenvalues 
of $\hH(\lambda)$.
Since this Hamiltonian becomes non-Hermitian for complex $\lambda$ values, the different energy levels 
can be considered as individual sheets of a Riemann surface that represents the one-to-many mapping between 
$\lambda$ and the quantised energy levels.\cite{Burton2019a,Marie2021}
Exceptional points (EPs) are the most common type of singularities on these Riemann surfaces. 
An EP is a square-root branch point that occurs when two energy levels become identical.\cite{Heiss2005,Heiss2012}
As a non-Hermitian (complex-symmetric) eigenvalue problem, two eigenfunctions $\Psi_+(\lambda)$ and $\Psi_-(\lambda)$  
coalesce at the EP such that $\Psi_+(\lambda) = \Psi_-(\lambda)$.\cite{MoiseyevBook}
When these points occur in the complex plane, an associated avoided crossing between two energy levels will 
exist on the real axis.
The sharpness of this avoided crossing is associated with the distance of the EP from the real axis, and 
thus a quantum phase transition will occur when an EP lies \textit{on} the real axis.\cite{Heiss2012,Sun2022}

Applying these concepts allows us to investigate whether excited-state perturbation series are convergent 
by assessing the complex-analytic properties of the $k^\text{th}$ sheet of the $\lambda$-dependent Riemann surface.
Locating EPs by inspection is difficult for anything beyond the simplest two- or three-level systems.
Instead, we use quadratic Pad\'{e} approximants to numerically estimate the position of the dominant 
singularity $\lep$.\cite{Goodson2012,Goodson2019,Mayer1985}
A quadratic approximant is defined as
\begin{equation}
	\label{eq:QuadApp}
	E_{[d_P/d_Q,d_R]}(\lambda) = \frac{1}{2 Q(\lambda)} \qty[ P(\lambda) \pm \sqrt{P^2(\lambda) - 4 Q(\lambda) R(\lambda)} ],
\end{equation}
with the polynomials 
\begin{align}
	\label{eq:PQR}
	P(\lambda) & = \sum_{j=0}^{d_P} p_j \lambda^j,
	&
	Q(\lambda) & = \sum_{j=0}^{d_Q} q_j \lambda^j, 
	&
	R(\lambda) & = \sum_{j=0}^{d_R} r_j \lambda^j.
\end{align}
This structure is designed to model the presence of square-root branch points in a function
and matches the truncated Taylor series expansion to $n^\text{th}$ order when $d_P + d_Q + d_R + 1 = n$.
Therefore, computing the $[d_P/d_Q,d_R]$ approximant requires the first $d_P + d_Q + d_R + 2$ terms 
in the corresponding Taylor series.
The approximate EPs are given by the roots of the polynomial 
$P^{2}(\lambda) - 4Q(\lambda) R(\lambda)$ and provide an estimate for the singularities
on the true MP energy function $E(\lambda)$.


\subsection{Computational details}
\label{subsec:compdet}
All HF calculations were performed using the standard SCF solver and the MOM extension
available in the \textsc{PySCF} package,\cite{PySCFb}
from which molecular orbital coefficients, and one- and two-electron integrals were computed.
MP$n$ calculations were evaluated using an in-house computational implementation represented 
in the full Hilbert space. 
Parameters for the quadratic approximants were computed by solving the corresponding system
of linear equations, as described in Ref.~\onlinecite{Fasondini2019}.
Due to the limitations of our computational implementation, all calculations were performed in the 
minimal STO-3G basis.\cite{Hehre1969}

\section{Results and Discussion}
\label{sec:results}

\subsection{Molecular \ce{H2}}
\label{subsec:H2groundstate}

\begin{figure}[htb]
\includegraphics[width=\linewidth]{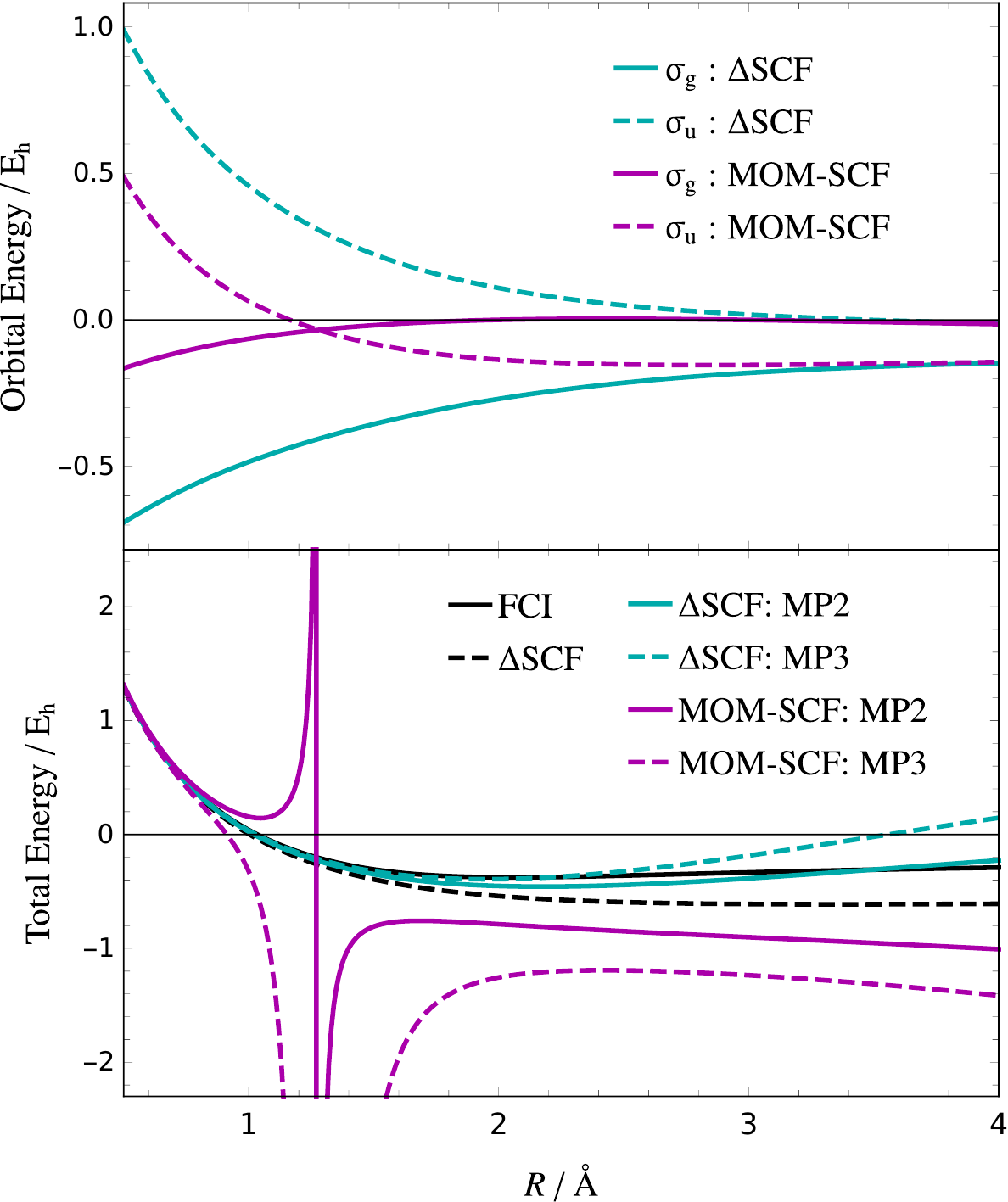}
\caption{Orbital energies for \ce{H2} using the $\Delta$SCF or MOM-SCF reference Fock matrices (top panel).
Comparison of the MP2 and MP3 total energies for the state using 
the $\Delta$SCF or MOM-SCF strategies (bottom panel).}
\label{fig:h2energies}
\end{figure}

We first consider molecular \ce{H2}, for which
it is well known that the ground-state MP$n$ series becomes divergent for large bond lengths
as a result of the near-degeneracy in the molecular orbitals.\cite{Gill1988}
The  closed-shell excitation in this system is the $2\,^1\Sigma^{+}_\text{g}$ state 
corresponding to the doubly-excited $\sigu^2$ configuration.
The only Hamiltonian coupling term is between this configuration and the $\sigg^2$ 
ground state, creating an effective two-level system.
The simplicity of this example ensures that all the energy surfaces can be plotted 
and visualised in the complex-$\lambda$ plane.

The $\Delta$SCF strategy employs the reference Hamiltonian using the orbitals and Fock matrix 
identified from a
ground-state Hartree--Fock solution, giving the two-sheeted Riemann surface in 
Fig.~\ref{fig:h2convergence}\textcolor{blue}{A}.
When the $\sigu^2$ configuration is used as the reference determinant, the MP perturbation
series corresponds to the Taylor series expansion evaluated on the upper sheet of this 
Riemann surface.
Only one complex-conjugate pair of EPs exists, connecting the ground and
excited $^1\Sigma^{+}_\text{g}$ states in a direct analogy to the two-site Hubbard model 
considered in Ref.~\onlinecite{Marie2021}.
Therefore, the radius of convergence for the excited-state perturbation expansion is identical 
to the ground-state approximation.
Numerically calculating $\rc$ using a $[2/2,2]$ quadratic approximant
(Fig.~\ref{fig:h2convergence}\textcolor{blue}{B}) reveals that divergent perturbation expansions
occur for $R > \SI{2.06}{\angstrom}$, as illustrated for two 
internuclear distances either side of this point in Fig.~\ref{fig:h2convergence}\textcolor{blue}{C}. 
The domain where the $\Delta$SCF perturbation series diverges is reflected in the regions where the 
MP3 total energy provides a poor approximation to the total excited-state energy (Fig.~\ref{fig:h2energies}, bottom panel).
Notably, the MP2  total energy remains a relatively good approximation up to $R = \SI{4}{\angstrom}$, 
although this is a numerical coincidence rather than a systematically improvable result.


\begin{figure}[htb]
\includegraphics[trim=0pt -10pt 0pt 0pt, clip, width=0.7\linewidth]{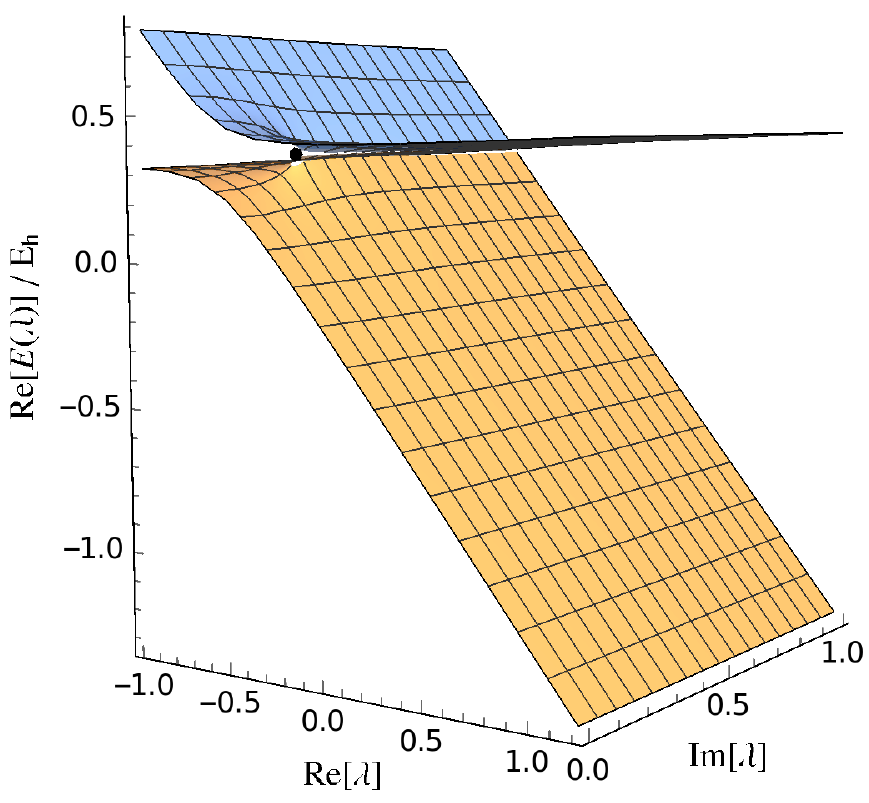}
\caption{Riemann surface representing the ground and excited states using the MOM-SCF excited reference state 
in \ce{H2} (STO-3G) at the equilibrium geometry $R(\ce{H-H}) = \SI{0.76}{\angstrom}$. 
The EP moves closer to the origin compared to the $\Delta$SCF reference state 
(Fig.~\ref{fig:h2convergence}\textcolor{blue}{A}).}
\label{fig:h2momRiemann}
\end{figure}

The MOM-SCF strategy optimises the orbitals for the excited 
$2\,^1\Sigma^{+}_\text{g}$ state and constructs the Fock matrix 
using the corresponding excited-state density.
Since both the reference state and Hamiltonian change, 
the energy surface $E(\lambda)$ is different to $\Delta$SCF (Fig.~\ref{fig:h2momRiemann}).
While we would expect the MOM procedure to improve the accuracy of the perturbation expansion, 
it actually leads to a divergent series  at almost every bond length.
Furthermore, the radius of convergence becomes zero (Fig.~\ref{fig:h2convergence}\textcolor{blue}{B})
around $R\approx\SI{1.28}{\angstrom}$, leading to a singularity 
in the binding curve (Fig.~\ref{fig:h2energies}, bottom panel).
While the dominant singularity moves continuously through the origin, its magnitude 
$\abs{\lambda}$ (and thus $\rc$) exhibits a cusp at $\lambda = 0$.
This divergence is particularly surprising because
the minimal basis set does not permit any orbital relaxation; the molecular orbitals are fixed by symmetry
and are the same in the ground and excited state.
Therefore, the only difference between the MOM-SCF and $\Delta$SCF strategies 
is the definition of the reference Fock matrix.

The failure of the MOM-SCF strategy can be understood by looking at the corresponding effect on 
the orbital energies (Fig.~\ref{fig:h2energies}, top panel), which determine the zeroth-order energy.
The anti-bonding $\sigu$ orbital is occupied in the excited state 
and provides the excited-state  density used to define the Fock matrix,
while the bonding $\sigg$ orbital is empty. 
This change in the reference density raises and lowers the $\sigg$ and $\sigu$ orbital energies, respectively.
An artificial crossing in the orbital energies occurs at $R=\SI{1.28}{\angstrom}$, at which point the
$\sigg^2$ and the $\sigu^2$ configurations become degenerate 
and $\rc = 0$ (Fig.~\ref{fig:h2convergence}\textcolor{blue}{B}).
Since the $\sigu$ orbital remains the lowest energy for all longer bond lengths, the $\sigu^2$ configuration
is the ground state of the reference Hamiltonian and no longer provides a suitable reference 
state for the excited-state perturbation series.
Consequently, defining the Fock matrix from the excited-state density, as occurs in the MOM-SCF approach,
leads to artificial changes in the energetic ordering of states that results  
in a divergent excited-state perturbation series.


\subsection{Closed-shell \ce{H2O} excitations}
\label{sec:H2O}

\begin{figure}[b]
\includegraphics[width=\linewidth]{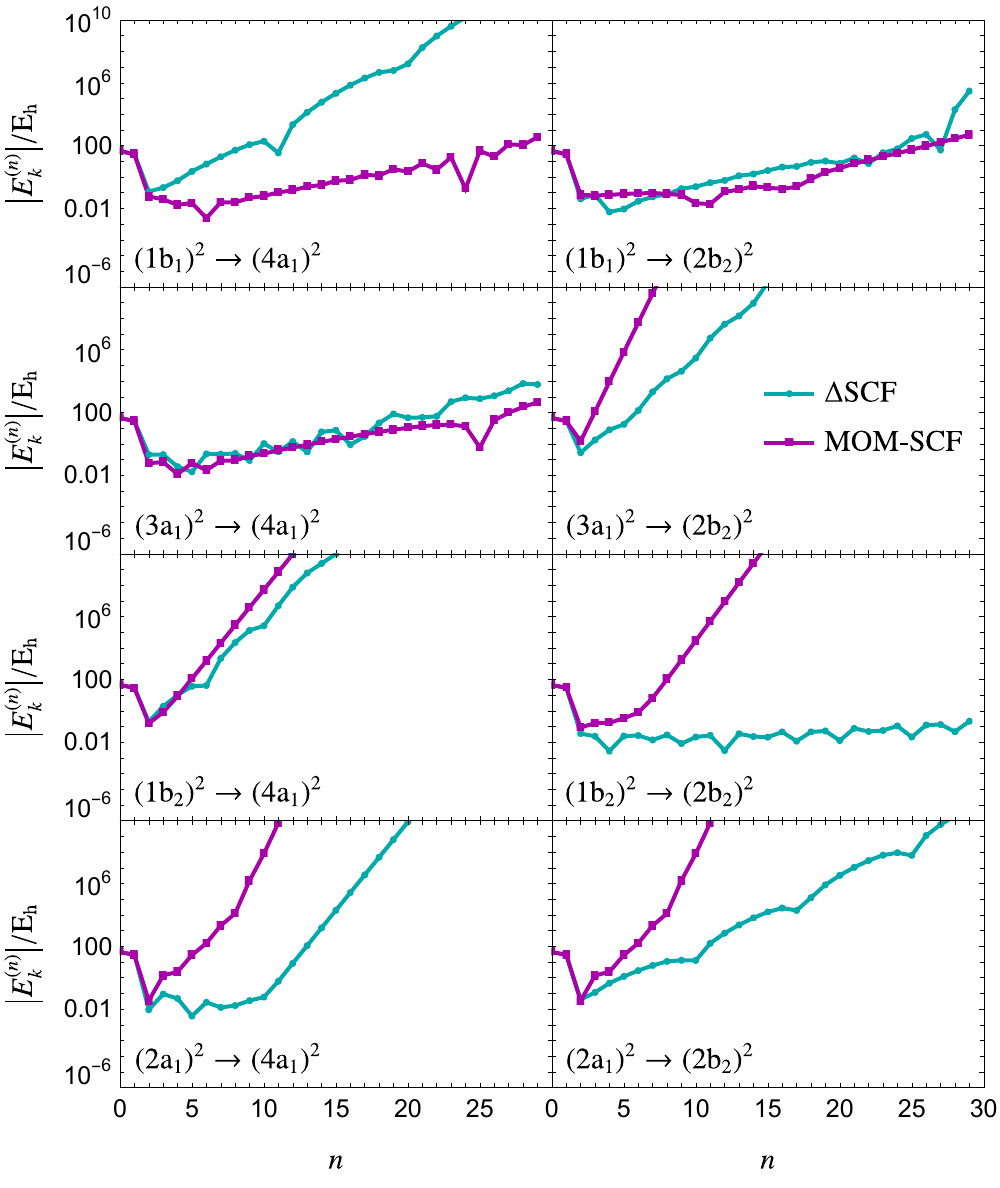}
\caption{Divergence of the excited-state MP series for the closed-shell valence double excitations 
in \ce{H2O} (STO-3G) at the equilibrium geometry.}
\label{fig:h2o_convergence}
\end{figure}

\begin{table*}[t]
	\caption{Analysis of the excited-state total energies for \ce{H2O} (STO-3G) at the ground state equilibrium geometry. 
	Quadratic approximants of order $[5/5,5]$ are used to identify the dominant singularity, and thus the radius of convergence.
    }
	\label{tab:h2o_res}
	\begin{ruledtabular}
	\begin{tabular}{ llccccc } 
		State & Method & MP1 / $\Eh$ & MP2 / $\Eh$ & MP3 / $\Eh$  & $\lep$ & $\rc$ \\
		\hline
		\multirow{2}{5em}{$\mathrm{(1b_1)^2 \rightarrow (4a_1)^2}$} & $\Delta$SCF & -73.771 & -73.880 & -73.678 & $\hphantom{-}0.250\pm0.096\,\mathrm{i}$ & 0.267 \\
		                                                                & MOM-SCF  & -73.849 & -73.907 & -73.944 & $\hphantom{-}0.547\pm0.012\,\mathrm{i}$ & 0.547 \\ \hline
		\multirow{2}{5em}{$\mathrm{(1b_1)^2 \rightarrow (2b_2)^2}$} & $\Delta$SCF & -73.648 & -73.688 & -73.615 & $\hphantom{-}0.500\pm0.000\,\mathrm{i}$ & 0.500 \\
		                                                                & MOM-SCF  & -73.682 & -73.754 & -73.821 & $\hphantom{-}0.561\pm0.007\,\mathrm{i}$ & 0.561 \\ \hline
		\multirow{2}{5em}{$\mathrm{(3a_1)^2 \rightarrow (4a_1)^2}$} & $\Delta$SCF & -73.525 & -73.733 & -73.940 & $           -0.440\pm0.324\,\mathrm{i}$ & 0.547 \\
		                                                                & MOM-SCF  & -73.614 & -73.673 & -73.742 & $           -0.600\pm0.105\,\mathrm{i}$ & 0.609 \\ \hline
		\multirow{2}{5em}{$\mathrm{(3a_1)^2 \rightarrow (2b_2)^2}$} & $\Delta$SCF & -73.465 & -73.743 & -71.966 & $           -0.082\pm0.081\,\mathrm{i}$ & 0.116 \\
		                                                     & MOM-SCF$^\text{\,a}$  & -73.488 & -74.976 &-190.630 & $\hphantom{-}0.011\pm0.003\,\mathrm{i}$ & 0.012 \\ \hline
		\multirow{2}{5em}{$\mathrm{(1b_2)^2 \rightarrow (4a_1)^2}$} & $\Delta$SCF & -73.430 & -73.228 & -71.314 & $\hphantom{-}0.083\pm0.072\,\mathrm{i}$ & 0.110 \\
		                                                                & MOM-SCF  & -73.430 & -73.588 & -74.396 & $\hphantom{-}0.071\pm0.008\,\mathrm{i}$ & 0.072 \\ \hline
		\multirow{2}{5em}{$\mathrm{(1b_2)^2 \rightarrow (2b_2)^2}$} & $\Delta$SCF & -73.128 & -73.093 & -73.116 & $           -0.314\pm0.792\,\mathrm{i}$ & 0.852 \\
		                                                                & MOM-SCF  & -73.133 & -73.220 & -73.387 & $\hphantom{-}0.059\pm0.000\,\mathrm{i}$ & 0.059 \\ \hline
		\multirow{2}{5em}{$\mathrm{(2a_1)^2 \rightarrow (4a_1)^2}$} & $\Delta$SCF & -72.143 & -72.152 & -72.243 & $\hphantom{-}0.074\pm0.000\,\mathrm{i}$ & 0.074 \\
		                                                                & MOM-SCF  & -72.154 & -72.144 & -72.934 & $\hphantom{-}0.083\pm0.000\,\mathrm{i}$ & 0.083 \\ \hline
		\multirow{2}{5em}{$\mathrm{(2a_1)^2 \rightarrow (2b_2)^2}$} & $\Delta$SCF & -71.967 & -71.927 & -72.044 & $\hphantom{-}0.290\pm0.130\,\mathrm{i}$ & 0.318 \\
		                                                                & MOM-SCF  & -71.968 & -72.003 & -73.398 & $           -0.003\pm0.000\,\mathrm{i}$ & 0.003 
    \end{tabular}
    \footnotesize a. A [2/2,2] approximant is used to identify $\lep$ for the MOM-SCF 
     $\mathrm{(3a_1)^2 \rightarrow (2b_2)^2}$ excitation as higher-order approximants could not be solved numerically.
	\end{ruledtabular}
\end{table*}

The excitations of \ce{H2O} are frequently used to test excited-state methods and are
generally considered to have Rydberg character that should favour the MOM-SCF strategy.
Here, we consider the closed-shell double valence excitations at the equilibrium geometry\cite{Attila2005} 
[$R(\ce{O-H}) = \SI{0.9572}{\AA}$; $\angle{\ce{HOH}} = \SI{104.5}{\degree}$].
These excitations correspond to transferring a pair of electrons from one of the occupied orbitals
($\mathrm{2a_1}$, $\mathrm{1b_2}$, $\mathrm{3a_1}$, $\mathrm{1b_1}$) into either the $\mathrm{4a_1}$ or 
$\mathrm{2b_2}$ virtual orbitals, giving eight excited states.
As each state has $^1\mathrm{A}_1$ symmetry and corresponds to a double excitation, 
they are all coupled with the ground state through the Hamiltonian.
Furthermore, each pair of excited states corresponding to an excitation from the same occupied
orbital or into the same virtual orbital are also coupled through the Hamiltonian.

The $\Delta$SCF strategy gives a divergent MP series for every excitation (Fig.~\ref{fig:h2o_convergence}), 
often at a very large rate.
Quadratic approximants of order $[5/5,5]$ were used to identify the dominant singularity for each excitation, 
providing an approximation to the radius of convergence (Table~\ref{tab:h2o_res}).
The most rapidly divergent expansions, corresponding to the $\mathrm{(3a_1)^2 \rightarrow (1b_2)^2}$, 
$\mathrm{(1b_2)^2 \rightarrow (4a_1)^2}$, and $\mathrm{(2a_1)^2 \rightarrow (4a_1)^2}$ excitations,  
have $\rc~\approx~0.1$ and give erratic low-order approximations to the total energy.
In these cases, the presence of an EP close to $\lambda = 0$ indicates a near degeneracy in 
the reference Hamiltonian, which is supported by the small differences between 
the total energies of the $\Delta$SCF reference determinants (given by the MP1 energy). 
Hence, the more complex orbital spectrum for \ce{H2O}
leads to more near-degeneracies between excited states that result in divergent series.

For the three lowest-energy excited-states considered, adding orbital optimisation through MOM-SCF
marginally increases the radius of convergence, but not enough to prevent the divergence at $\lambda = 1$.
However, the same approach significantly worsens the divergence for the remaining five excitations, giving 
$\rc < 0.1$ with an EP very close to $\lambda = 0$.
For the $\mathrm{(3a_1)^2 \rightarrow (1b_2)^2}$ with $\rc~\approx~0.01$, we experience numerical 
instabilities in the quadratic approximants at higher orders, and thus we are limited to the $[2/2,2]$ approximant.
These data support the conclusion from the \ce{H2} excited state: the MOM-SCF 
reference can reduce the energy gap between the newly occupied and unoccupied orbitals involved in the excitation, 
increasing the severity of near degeneracies between the reference configurations.
In the worst case, the $\mathrm{(3a_1)^2 \rightarrow (2b_2)^2}$ excitation, the divergence is so rapid that 
even the MP3 total energy is completely meaningless (Table~\ref{tab:h2o_res}).

\begin{figure*}[htb]
\includegraphics[width=\linewidth]{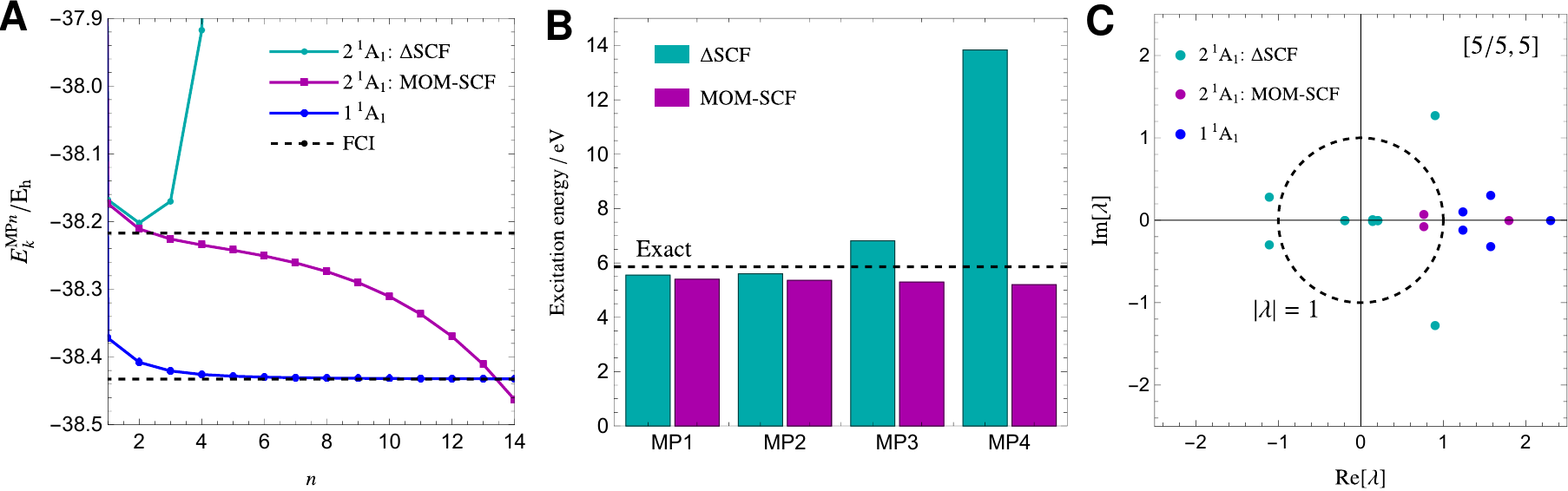}
\caption{%
(\textbf{A}) Both $\Delta$SCF and MOM-SCF give a divergent series for the 
$2\,^{1}\text{A}_1$ excited state of \ce{CH2} (STO-3G) at the singlet ground state geometry.
(\textbf{B}) State-specific orbital optimisation provides more consistent 
low-order approximations to the excitation energy.
(\textbf{C}) The $[5/5,5]$ quadratic approximant reveals that orbital optimisation
shifts the dominant singularities away from the origin in the complex-$\lambda$ plane.
 }
\label{fig:ch2_convergence}
\end{figure*}

From a qualitative perspective, the eight excited states considered are obtained by exciting electrons 
from orbitals of decreasing energy, and the total energies are expected to increase in the same order.
The $\Delta$SCF reference energies preserve this ordering, but it is lost at higher
orders of the perturbation series. 
Furthermore, while the total energy of a MOM-SCF reference state is always lower than the corresponding 
$\Delta$SCF configurations, as expected with orbital optimisation, this ordering is not necessarily
preserved for the MP2 and MP3 energies.
The erratic behaviour of these low-order approximations, and the high density of exact excited states,
complicates the physical interpretation and assignment to exact excitations.
Consequently, it is very difficult to extract reliable physical excitation energies for \ce{H2O} 
from low-order perturbation theory using excited-state reference determinants.

\subsection{Singlet excitation in methylene}
\label{sec:CH2}
Finally, we consider the methylene \ce{CH2} molecule.
The ground state has triplet $\mathrm{^{3}B_2}$ symmetry due to the small orbital energy gap between the 
in-plane $\mathrm{3a_1}$ (carbon $\mathrm{sp^2}$) and out-of-plane $\mathrm{1b_2}$ (carbon $\mathrm{2p_y}$) orbitals.
The two lowest-energy closed-shell states correspond to the 
$1\,^{1}\mathrm{A}_1$ ground state and the $2\,^{1}\mathrm{A}_1$ doubly-excited state.
We consider the vertical excitation at the singlet equilibrium 
geometry [$R(\ce{C-H}) = \SI{1.111}{\AA}$; $\angle{\ce{HCH}} = \SI{102.4}{\degree}$] 
given in Ref.~\onlinecite{Minaev2002}.
Previous calculations by Lee \etal{}\ have shown that orbital optimisation can significantly improve 
the accuracy of this excitation energy computed using MP2,\cite{Lee2019b}
although the series convergence was not considered.

The ground-state $1\,^{1}\text{A}_1$ series converges rapidly, with the MP5 approximant
providing an accuracy of $4\,\mathrm{mE_h}$ (Fig.~\ref{fig:ch2_convergence}\textcolor{blue}{A}).
As the $2\,^{1}\mathrm{A}_1$ is the lowest excited state with this symmetry, we expect it to be
less affected by the near-degeneracy issues seen in the \ce{H2O} excitations.
However, the $2\,^{1}\text{A}_1$ excited-state energies still diverge for both the $\Delta$SCF and 
the MOM-SCF reference states, although the MOM-SCF series diverges at a slower rate.
The corresponding MP2 excitation energies are remarkably accurate for both cases,
although the $\Delta$SCF excitation energies deteriorate rapidly beyond MP3
(Fig.~\ref{fig:ch2_convergence}\textcolor{blue}{B}).
However, the slower divergence in the MOM-SCF excitation energy means that the lack of systematic
improvability is less noticeable at low orders, obscuring the reliability of 
practical calculations.

In contrast to \ce{H2O}, the MOM-SCF perturbation series in \ce{CH2} diverges at a slower rate than $\Delta$SCF.
This convergence behaviour is elucidated by the singularity structure of the $\lambda$-dependent Hamiltonian
in the complex-$\lambda$ plane (Fig.~\ref{fig:ch2_convergence}\textcolor{blue}{C}).
The $\Delta$SCF series shows fast and erratic divergence due to singularities close to 
the origin. 
Orbital optimisation using MOM-SCF shifts these singularities away from the origin along the real-$\lambda$
axis, although they remain within the unit circle and the series still diverges  at $\lambda = 1$
(with a slower rate).
We believe that the improved performance of MOM-SCF for this system arises because we are 
targeting the lowest excited state with $^{1}\text{A}_1$ symmetry. 
Therefore, lowering the energy of the occupied excited orbitals through orbital optimisation 
increases the energy gap to the higher excited states with this symmetry, while the energy
gap to the ground state remains sufficiently large.
Consequently, this orbital optimisation reduces the 
severity of near-degeneracies with other excited states, leading to a more convergent series.


\begin{table}[b]
\caption{The $[n/n,n]$ sequence of quadratic approximants gives more accurate 
total energies for the $2\,\mathrm{^1A_2}$ excited state in \ce{CH2} (STO-3G).
Energies are given in $\mathrm{E_h}$.}
\label{tab:ch2_approx}
	\begin{ruledtabular}
	\begin{tabular}{ lccc }
   \multirow{2}{*}{Order} & \multirow{2}{*}{$1\,^{1}\mathrm{A_1}$} & \multicolumn{2}{c}{$2\,^{1}\mathrm{A_1}$} \\
                                                                     \cline{3-4}
                          &                                        & $\Delta\text{SCF}$ & MOM-SCF \\
    \hline
    $[1/1,1]$ & $-38.429\,641$ & $-38.206\,797$ & $-38.182\,250$ \\
    $[2/2,2]$ & $-38.432\,314$ & $-38.193\,280$ & $-38.217\,375$  \\
    $[3/3,3]$ & $-38.432\,546$ & $-38.215\,900$ & $-38.217\,005$ \\
    $[4/4,4]$ & $-38.432\,558$ & $-38.218\,599$ & $-38.217\,094$ \\
    $[5/5,5]$ & $-38.432\,495$ & $-38.218\,620$ & $-38.217\,098$ \\
    \hline
    Exact & $-38.432\,495$ & $-38.217\,100$ & $-38.217\,100$ 
    \end{tabular}
	\end{ruledtabular}
\end{table}

In addition to elucidating the singularity structure, quadratic approximants to the MP energy 
function $E(\lambda)$ can provide more accurate approximations to the physical energy 
at $\lambda = 1$.\cite{Goodson2012,Goodson2019}
By modelling the square-root singularity structure of $E(\lambda)$, the sequence of $[n/n,n$] 
approximants with increasing order can turn a divergent MP expansion into a convergent series.\cite{Goodson2000}
Table~\ref{tab:ch2_approx} suggests that these approximants give a rapidly 
converging series for the ground state and MOM-SCF excited-state energies evaluated at $\lambda = 1$, 
giving an error within $0.002\,\mathrm{mE_h}$ for the $[5/5,5]$ term.
The $\Delta$SCF approximants are more stable than the MP series, but are less accurate than MOM-SCF.
Consequently, while orbital optimisation does not turn the divergent excited-state
series into a convergent one, the transformed MP energy function is better suited to 
low-order quadratic approximants that give very accurate total energies.

\section{Concluding Remarks}
\label{sec:conclusions}

Second-order MP perturbation theory is widely used as the first approximation to the correlation energy 
due to its relatively modest computational scaling.
However, systematic improvability of the MP perturbation theory relies on the series convergence,
which requires states that are accurately described by a single Slater determinant 
and a lack of near-degeneracies in the reference configurations.
We have investigated the convergence behaviour of the single-determinant MP series for 
stereotypical closed-shell doubly-excited states
using higher-energy reference Slater determinants, both with and without orbital optimisation.
Our results suggest that excited reference states are very unlikely to give a convergent perturbation series.
While low-order approximations can give physically meaningful energies for some states, others 
show such rapid divergence that even MP3 is unreliable.

Orbital optimisation of the reference state, using methods such as MOM-SCF,\cite{Gilbert2008} 
generally lowers the energies of the occupied orbitals involved in the excitation 
and the total energy of the Slater determinant.
This optimisation, combined with an alternative reference Fock operator built using the excited-state density, 
can either improve or worsen the divergence of the perturbation series. 
When the target excited state lies in the middle of a dense region of the excited-state spectrum, orbital optimisation
appears to worsen the degeneracies that can cause very rapid series divergence. 
On the contrary, when the target state is lower than other excitations with the same symmetry, 
orbital optimisation can increase the energetic gap to other excited states and reduce the rate of divergence.
It is therefore feasible that orbital optimisation may turn a divergent $\Delta$SCF series into a convergent 
one, although we have not found any examples with this behaviour, and it may be difficult to
predict which excited states are most suitable for MP theory. 
Furthermore, even if orbital optimisation does not give a convergent series, it can still provide more accurate total
energies using resummation techniques such as quadratic approximants, although the lack of a corresponding 
wave function means that other system properties cannot be computed through these approaches.
A variety of different resummation techniques have been developed to obtain meaningful results for ground-state 
calculations,\cite{Mihalka2017,Mihalka2019,Surjan2018,Cizek1993,Goodson2000,Marie2021,Goodson2000a,Cizek1996}
which may improve the accuracy of excited state-specific perturbation theory.

The current study only considered small basis sets and closed-shell excitations. 
While increasing the basis set may increase the flexibility of the excited-state optimisation, 
and enlarge the radius of convergence, it is also well known that adding more diffuse functions can 
lead to divergences associated with M\o{}ller--Plesset critical points.\cite{Olsen2000,Sergeev2005,Sergeev2006}
It remains to be seen whether these effects will further exacerbate the divergences of excited-state
perturbation expansions.
On the other hand, extending the study to open-shell excitations would require minimal 
multiconfigurational reference states to account for the degenerate configurations and avoid spin contamination
(e.g.\ Refs.~\onlinecite{Kossoski2022a,Shea2020,Zhao2020a,Hardikar2020,Shea2018}).
Subsequent low-order perturbation theory can provide more accurate excitation energies, but
intruder state problems are still observed in some cases.\cite{Kossoski2022a,Clune2020}
Without rigorous testing, it is hard to predict whether these multiconfigurational perturbation
expansions will converge, although one might expect the lowest state of each symmetry to behave in a 
similar way to ground-state perturbation theory.

The use of higher-energy SCF solutions as reference states for perturbation theory has proven to be
a highly delicate task, and is likely to become more difficult in larger systems with a higher spectral 
density of excited states.
Since these divergent expansions result from near degeneracies in the excited states, it is unlikely that 
this situation can be improved using an alternative Hamiltonian partitioning, or with level-shifted approaches.
Instead, we believe that research into post-mean-field correlation corrections for excited state-specific Slater 
determinants should focus on non-perturbative methods such as coupled cluster theory, for which 
recent studies have shown very promising results.\cite{Marie2021a,Kossoski2021,Mayhall2010,Lee2019b}
Alternatively, many excited states fundamentally require a multi-configurational state-specific 
approach.\cite{Olsen1983,Shea2018,Tran2019,Tran2020,Marie2023}
Second-order perturbation corrections to complete-active-space methods\cite{Andersson1990,Andersson1992} 
(CAS-PT2) or nonorthogonal configuration interaction\cite{Burton2020} (NOCI-PT2) remain the 
most promising candidates for capturing dynamic correlation, although these methods often require empirical
level shifts or the IPEA approach to avoid divergences.

\section*{Acknowledgements}
H.G.A.B.\ was supported by New College, Oxford through the Astor Junior Research Fellowship.
The authors thank Antoine Marie for useful discussions, and David Tew for support and computing resources.

\section*{Data availability}
The data that supports the findings of this study are available within the article.

\section*{References}
\bibliography{manuscript}

\end{document}